\documentclass[showpacs,amssymb]{revtex4}
\usepackage{graphicx}
\usepackage{bm}
\begin{document}  

\title{
Accurate and realistic initial data for black hole-neutron star binaries.
}  
\author{Philippe Grandcl\'ement}
\email{philippe.grandclement@obspm.fr}
\affiliation{
Laboratoire de l'Univers et de ses Th\'eories, UMR 8102 du C.N.R.S.,
Observatoire de Paris, F-92195 Meudon Cedex, France}

\date{May18, 2006.}

\def\l{\left}
\def\r{\right}
\def\pa{\partial}
\def\be{\begin{equation}}
\def\ee{\end{equation}}
\def\bea{\begin{eqnarray}}
\def\eea{\end{eqnarray}}

\begin{abstract} 
This paper is devoted to the computation of compact binaries composed of one black hole and one neutron star. The objects are assumed to be on exact circular orbits. Standard 3+1 decomposition of Einstein equations is performed and the conformal flatness approximation is used. The obtained system of elliptic equations is solved by means of multi-domain spectral methods. Results are compared with previous work both in the high mass ratio limit and for one neutron star with very low compactness parameter. The accuracy of the present code is shown to be greater than with previous codes. Moreover, for the first time, some sequences containing one neutron star of realistic compactness are presented and discussed.
\end{abstract} 
 
\maketitle

\section{Introduction}
\label{s:intro} 
Motivated by the various gravitational wave detectors coming online \cite{ligo,virgo}, numerical simulations of binary compact objects have been extensively investigated in the last years. With progress made for the evolution of both binary neutron stars (BNS) \cite{ShibaT06} and binary black holes (BBH) \cite{Preto06,BakerCCKM06,CampaLMZ06}, it seems timely to turn to the last interesting type of binary: a system composed of one black hole and one neutron star (BHNS). The evolution synthesis codes indicate that the detection rate of BHNS with LIGO/Virgo will be as high (if not higher) than for BNS systems (see Table 6 of \cite{BelczKB01}). Simulations of coalescing binary systems usually proceed in two steps. First, one needs to produce initial data that verify the Einstein constraint equations and that are as physically relevant as possible. Then, the initial configurations are evolved forward in time. Those steps involve rather different techniques and are equally challenging.

Most of the initial data for coalescing binaries rely on the quasiequilibrium hypothesis that assumes that the objects are on exact closed circular orbits. This is of course only an approximation because no closed orbits can exist for those systems in general relativity. However, this description should be rather accurate, at least for large separations. The quasiequilibrium approximation has been applied to BNS \cite{BaumgCSST98a,BaumgCSST98b,UsuiUE00,UryuE00,GourgGTMB01} and BBH systems \cite{PfeifTC00,GourgGB02,GrandGB02,CaudiCGP06}. In the BNS case, the evolution of such initial data have exhibited circular-like trajectories \cite{ShibaT06}. The long term evolution of the data in the BBH case is still underway. The aim of this paper is to apply the same technique to a BHNS, without assuming extreme mass ratios like in \cite{TanigBFS05}. On the contrary, a moderate mass ratio of 5 is used. This value is chosen mainly for comparisons with previous work \cite{TanigBFS06} but is sufficient to demonstrate the ability of the code to handle binaries with close masses. Let us mention that the simulations of the formation of compact binaries indicate a slightly lower mass ratio to be more probable \cite{BulikGB04}. The present work shares some properties with \cite{TanigBFS06} even if some details are different. Moreover, one of the main shortcomings of \cite{TanigBFS06} is the fact that only a NS of unrealistic compactness, as small as $0.0879$, is considered. This implies a mass close to $0.7 \, M_\odot$ for most of the available EOS, which is much smaller than the canonical value of $1.2-1.4 \, M_\odot$. On the contrary, in this work and for the first time, a BHNS with a realistic NS is computed. Indeed, the most compact star presented here has a compactness of $0.15$ which makes its mass in the range $1.2-1.5 \, M_\odot$ depending on the EOS.

\section{Equations}
\label{s:eq}
The standard 3+1 decomposition of the Einstein equations is used, in which the spacetime is foliated by a family of spatial hypersurfaces. The 4-dimensional metric is given in terms of the lapse function $N$, the shift vector $\beta^i$ and the spatial metric $\gamma_{ij}$. The evolution of the spatial metric is described by the second fundamental form: the extrinsic curvature tensor $K_{ij}$ which is the Lie derivative of $\gamma_{ij}$ along the normal to the hypersurfaces.

The conformal factor $\Psi$ is defined by $\gamma_{ij} = \Psi^4 \tilde{\gamma}_{ij}$ and by demanding that the determinant of the conformal metric $\tilde{\gamma}_{ij}$ is $1$ (so that the determinant of $\gamma_{ij}$ is $\Psi^{12}$). A similar conformal decomposition is also performed on the extrinsic curvature tensor. 

Before solving the constraint equations, there are some quantities that must be chosen. Those are the so-called ``freely specifiable'' variables. By this, one means that a solution of the constraint equations can be found for every choice of those variables. In the standard ``thin-sandwich'' approach, these variables are: 
the trace of the extrinsic curvature $K$ and its time derivative $\partial_t K$, the conformal metric $\tilde{\gamma}_{ij}$ and its time derivative 
$\tilde{u}_{ij} = \partial_t \tilde{\gamma}_{ij}$. By working in the corotating frame, the quasiequilibrium approximation amounts to neglecting the variations with time of all the quantities. In particular, one sets $\partial_t K = 0$ and $\tilde{u}_{ij}=0$. Maximum slicing $K=0$ and conformal flatness 
$\tilde\gamma_{ij}=f_{ij}$, $f_{ij}$ being the 3-dimensional flat metric, are also assumed. Even if the spatial metric for a binary system can not be flat, this approximation has proven to be more accurate that one could expect for both BBH and BNS \cite{DamouGG02,Blanc02,UryuLFGS05}. Let us note that the choices for $K$ and $\tilde{\gamma}_{ij}$ are different from the ones made in \cite{TanigBFS06} where the authors used values derived from the Kerr-Schild metric.

The mathematical problem is then to solve the following set of five coupled elliptic equations for the two scalars $\Psi$ and $N$ and the vector field $\beta^i$~:
\bea
\label{e:lapse}
\Delta N &=& 4\pi N\Psi^4 \l(E+S\r) + N\Psi^4 \tilde{A}_{ij}\tilde{A}^{ij} -2\bar{D}_i\ln\Psi\bar{D}^iN \\
\label{e:psi}
\Delta \Psi &=& -2\pi\Psi^5 E - \frac{\Psi^5}{8} \tilde{A}_{ij}\tilde{A}^{ij} \\
\label{e:beta}
\Delta \beta^i + \frac{1}{3} \bar{D}^i\bar{D}_j\beta^j &=& 16 \pi N\Psi^4 \l(E+p\r) U^i + 2\tilde{A}^{ij} \l(\bar{D}_jN - 6N \bar{D}_j\ln\Psi\r)
\eea
where all the operators are associated with the flat metric. The conformal extrinsic curvature tensor $\tilde{A}^{ij} = \Psi^4 K^{ij}$ relates to the lapse and shift vector by $\tilde{A}^{ij} = 1/2N \l(\bar{D}^i \beta^j + \bar{D}^j \beta^i - 2/3 \bar{D}_k\beta^k f^{ij}\r)$.
$E$, $S$, $p$ and $U^i$ are matter terms describing the NS fluid: $p$ is the pressure, and $E$ and $S$ are the matter energy density and the trace of the stress energy tensor, both measured by the Eulerian observer (whose 4-velocity coincides with the normal to the spatial hypersurfaces). $U^i$ is the fluid 3-velocity with respect to the Eulerian observer. Explicit expressions for all those matter terms can be found, for example, in \cite{GourgGTMB01}. In this paper, a polytropic equation of state relates the pressure to the fluid baryon number density by $p = \kappa n^\Gamma$. In all the following $\Gamma$ is fixed to $2$ and $\kappa$ is varied to construct NS with various compactness. The star is also assumed to be irrotational. It implies that the velocity $U^i$ relates to a potential $\Phi$ via :
\be
U^i = \frac{1}{\Psi^4 \Gamma_n h}\bar{D}^i\Phi,
\ee
where $h$ is the fluid specific enthalpy and $\Gamma_n$ the Lorentz factor between the Eulerian observer and the fluid 
(see \cite{GourgGTMB01} for explicit expressions). The potential obeys the following equation :
\be\label{e:fluid}
\xi H \Delta \Phi + \bar{D^i}H\bar{D_i}\Phi = \Psi^4h\Gamma_n U_0^i\bar{D}_iH + \xi H\l(\bar{D}^i\Phi\bar{D}_i\l(H - \beta\r)
+\Psi^4hU_0^i\bar{D}_i\Gamma_n\r),
\ee
where $H = \ln h$, $\xi=d\ln H/ d\ln n$, $\beta=\ln\l(\Psi^2N\r)$ and $U_0^i$ is the 3-velocity of the corotating observer. The irrotational fluid also admits an integral of motion which can be expressed as :
\be
\label{e:integrale}
h N \frac{\Gamma}{\Gamma_0} = {\rm const.},
\ee
where $\Gamma$ is the Lorentz factor between the corotating observer and the fluid and $\Gamma_0$ between the corotating and the Eulerian observers.

The black hole is described in the framework of the apparent horizons, where non-trivial boundary conditions are imposed on a sphere. (see the work of \cite{CookP04} or \cite{GourgJ06} for a review). This idea has already been successfully applied to BBH systems (see \cite{CaudiCGP06} for the last application to date). However, in this work, like in \cite{GourgGB02, GrandGB02}, the lapse is set to 0 on the horizon. In doing so one needs to make a small correction on the shift vector to get a regular extrinsic curvature tensor on the horizon. In the case of the BBH it has been shown that this correction was small enough \cite{GrandGB02}. In the BHNS case, the effect of the neutron star on the horizon is even smaller, making this relative correction very small (the relative difference between the original and the corrected shift is at most $2\cdot 10^{-4}$ for the innermost configurations). An important difference with respect to what was done in \cite{GrandGB02} concerns the rotation state of the black hole. Indeed, it seems unlikely that it will be synchronized with the orbital motion and an irrotational black hole will be considered. Like in Sec. VB of \cite{CaudiCGP06}, a local rotation rate is imposed. This is done by imposing that, on the horizon, the corotating shift is $\beta^i = \Omega_r \l(\partial/\partial \varphi\r)^i$, the angle $\varphi$ being associated with the horizon. In the case of the corotating black holes of \cite{GrandGB02}, $\Omega_r$ is zero. When considering irrotational black holes, $\Omega_r$ must be determined to ensure that the quasi-local spin of the black hole vanishes. The appropriate value of $\Omega_r$ must be obtained numerically.
This is to be contrasted with what is done in \cite{TanigBFS06} where the authors impose irrotationality only to the first order (see Sec. VA of \cite{CaudiCGP06}), i.e. by demanding that $\Omega_r$ coincides with the orbital velocity $\Omega_0$. As will be seen later, the correction to this, measured by the ratio 
$f_r = \Omega_r/\Omega_0$ is of the order of $10\%$. $f_r$ is one when the first order is valid, that is when mass ratio is infinite, and smaller than unity when the masses become comparable.

Standard asymptotic flatness is used to get appropriate boundary conditions at infinity. 

\section{Numerics, tests and comparisons}

The system (\ref{e:lapse}-\ref{e:beta}) is solved using the LORENE library \cite{lorene}. This library is developed mainly at the Meudon site of Paris observatory and has been successfully applied to the 
computation of various problems in general relativity (see references at \cite{lorene}). The basic features of Lorene are the following: spectral methods in spherical coordinates are employed, mainly using spherical harmonics or trigonometrical functions for the angles $\l(\theta, \varphi\r)$ and Chebyshev polynomials for the radial coordinate $r$. Space is decomposed in various spherical-like shells, the spectral expansion being done in each of those domains. Space is compactified by means of the 
variable $u=1/r$ in the outermost domain. This enables one to impose exact boundary conditions at infinity, the computational domain covering the whole space. Two sets of such domains are used, one centered around each compact object and the equations (\ref{e:lapse}-\ref{e:beta}) are split on those two sets of domains (see for example 
\cite{GourgGTMB01} and \cite{GrandGB02} for explicit implementations of this method). For the NS, the first domain is adapted to match the surface of the star, thus getting rid of any Gibbs phenomenon that would be caused by discontinuities at the surface. With such methods, solving elliptic equations amounts to inverting some matrices. Explicit examples are given in \cite{GrandBGM01} for Poisson equations and the algorithm used to solve Eq. (\ref{e:fluid}) can be found in \cite{GourgGTMB01}. Let us mention that the code of  \cite{TanigBFS06} is also based the LORENE library. However the two codes have been written completely independently and the details of the two implementations are not the same.

The system (\ref{e:lapse}-\ref{e:beta}), along with the integral (\ref{e:integrale}) and the equation (\ref{e:fluid}) for the potential $\Phi$, is solved by iteration, until the fields converge to a given threshold (typically $10^{-7}$). During the course of the iteration, some quantities are changed in order to fulfill various requirements. For example, the central enthalpy of the NS is rescaled at each step in order to get a neutron star of given baryon mass (see Sec. IVD3 of \cite{GourgGTMB01}). The same technique is also used to modify the radius of the black hole to get a given irreducible mass, the position of the rotation axis to ensure that the total linear momentum vanishes and the local rotation rate of the black hole to impose irrotationality. The distance between the two objects can also be modified by these means. As in \cite{GourgGTMB01}, the orbital velocity is determined so that the zero of the gradient of enthalpy lies at the origin of the grid associated with the NS. 

For both objects, space is decomposed on typically 8 or 9 domains, each of them being covered by ${\rm N}_r\times{\rm N}_\theta\times{\rm N}_\varphi = 33 \times 21 \times 20$ points. The achieved precision can be measured by some global checks. For instance, if Eq. (\ref{e:psi}) is fulfilled, the ADM mass can be computed in two ways: either by the standard integral at infinity, or by a volume integral plus an integral on the BH horizon (see Sec. IIID of \cite{GourgGTMB01}). For all our configurations, the 
relative difference between the two values is less than a few times $10^{-5}$. Other criteria of this kind can be derived that test the other equations 
(using the total angular momentum $J$ and the generalized Smarr formula ; see Sec. IIIC and IIID of \cite{GourgGB02}). However, in those cases, the precision is also limited by the regularization of the shift. As already stated, the relative difference between the original and the corrected shift is at most a few times $10^{-4}$ which implies a relative difference of $10^{-2}$ between the two expressions of $J$ (analogues of Eqs.(67) and (68) of \cite{GrandGB02}). The two orders of magnitude between the regularization of the shift and the error on the global quantity $J$ is similar to what is observed in the BBH case (see Fig. 10 of \cite{GrandGB02}).

\begin{figure}
\includegraphics[height=6.5cm]{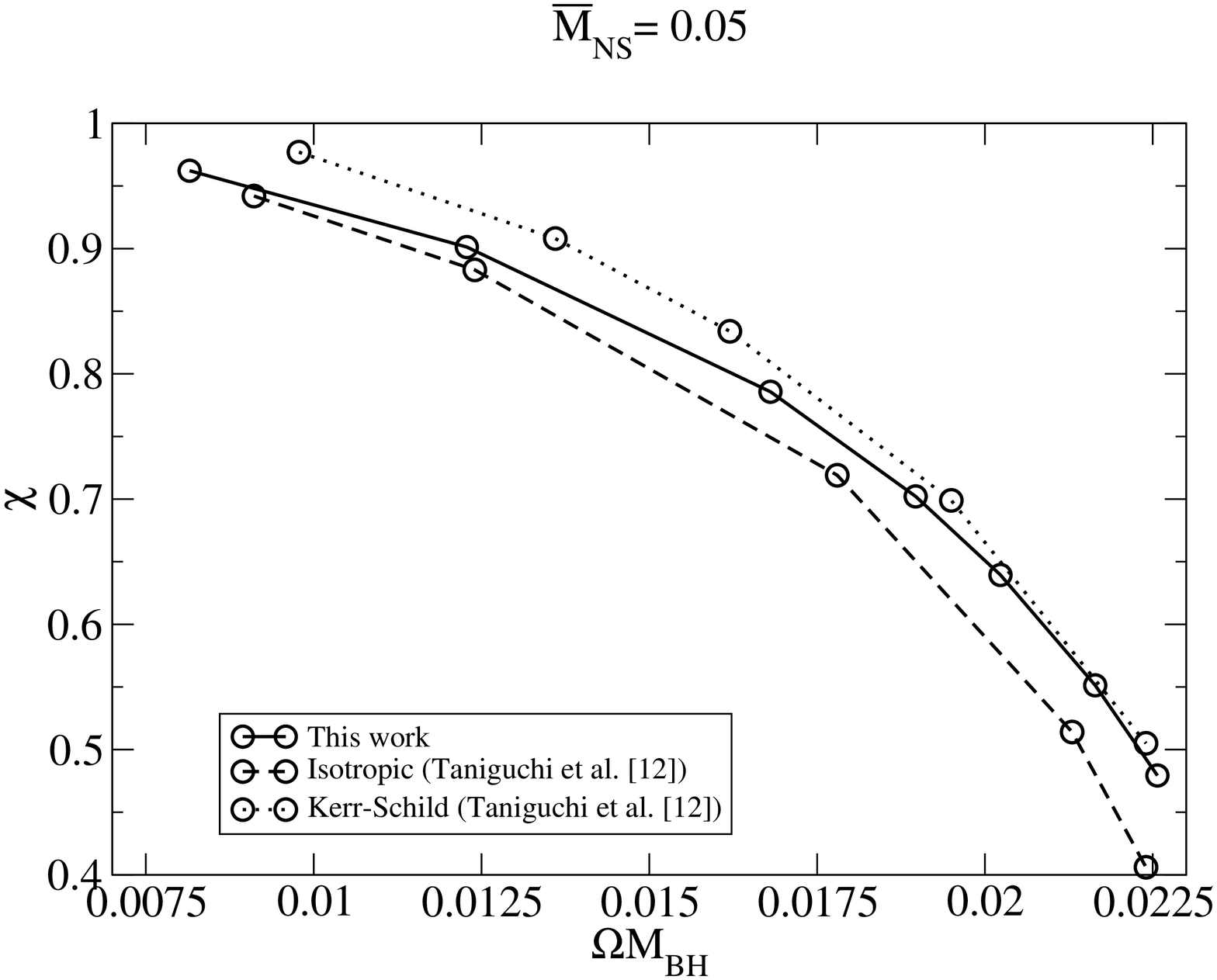}
\includegraphics[height=6.5cm]{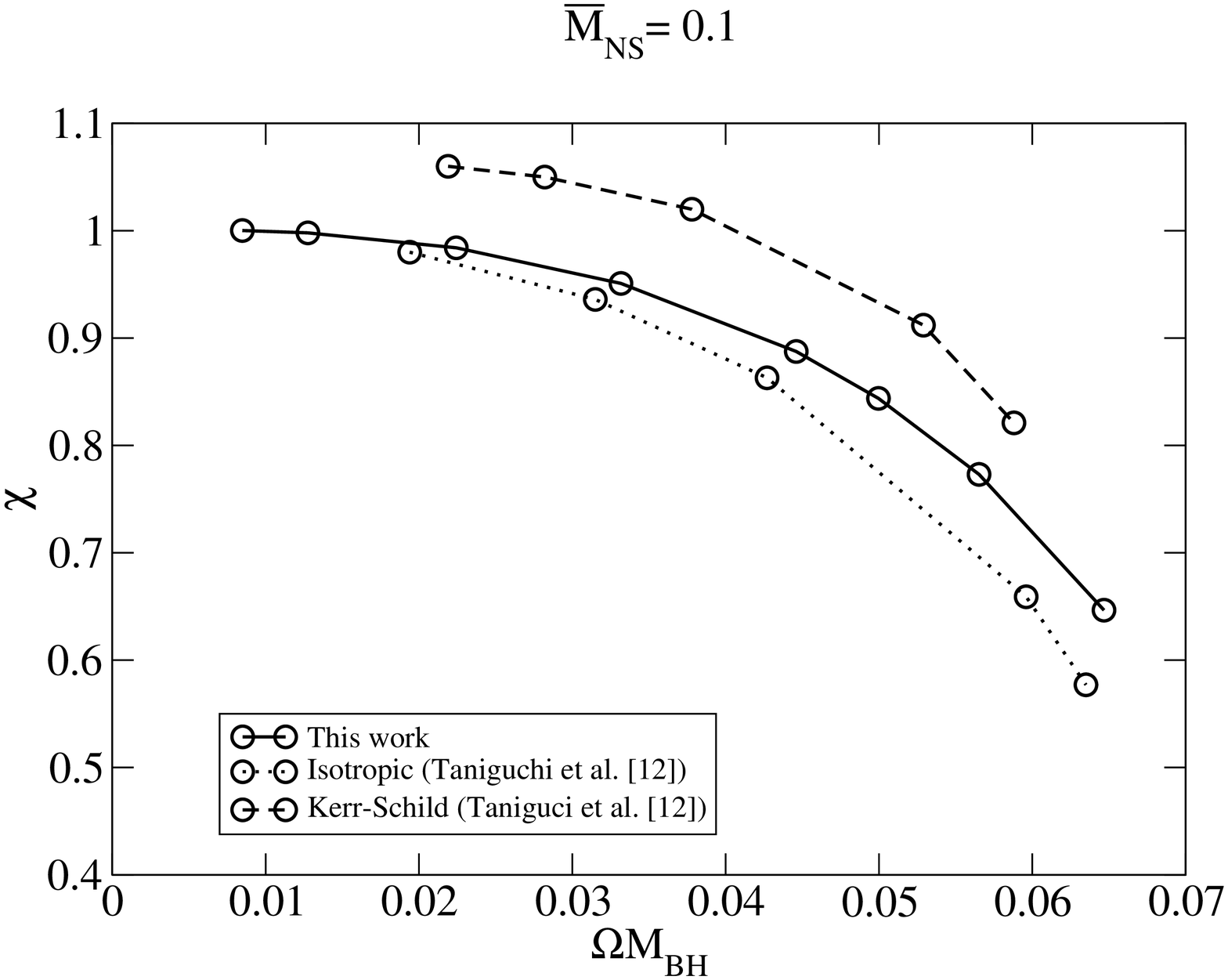}
\caption{\label{f:rapport_10}
Deformation indicator $\chi$ as a function of the orbital frequency for various approaches and $\bar{M}_{\rm NS} = 0.05$ (first panel) and 
$\bar{M}_{\rm NS}=0.01$ (second panel). The mass ratio is 10.}
\end{figure}

Another test is provided by comparing the results from this work with the ones obtained in the extreme mass ratio in \cite{TanigBFS05}. More precisely, in Fig. 
\ref{f:rapport_10}, the value of $\chi$ is shown as a function of the orbital frequency, for a mass ratio $M^{\rm irr}_{\rm BH}/M^{\rm b}_{\rm NS} = 10$. $M^{\rm irr}_{\rm BH}$ denotes the irreducible mass of the BH and $M^{\rm b}_{\rm NS}$ the baryonic mass of the NS, both quantities being kept constant along a sequence. $\chi$ is a measure of the deformation of the star. It is one for a spherical star and zero at the mass-shedding limit (see Eq. (52) of \cite{TanigBFS05} for the precise definition). The first panel of Fig. \ref{f:rapport_10} shows the value of $\chi$ for $\bar{M}^{\rm b}_{\rm NS} = 0.05$ and the second one for 
$\bar{M}^{\rm b}_{\rm NS} = 0.01$, the masses being expressed in standard polytropic units. The agreement with \cite{TanigBFS05} is good, especially for the isotropic background. The difference is more important with the Kerr-Schild approach.

\begin{figure}
\includegraphics[height=6.5cm]{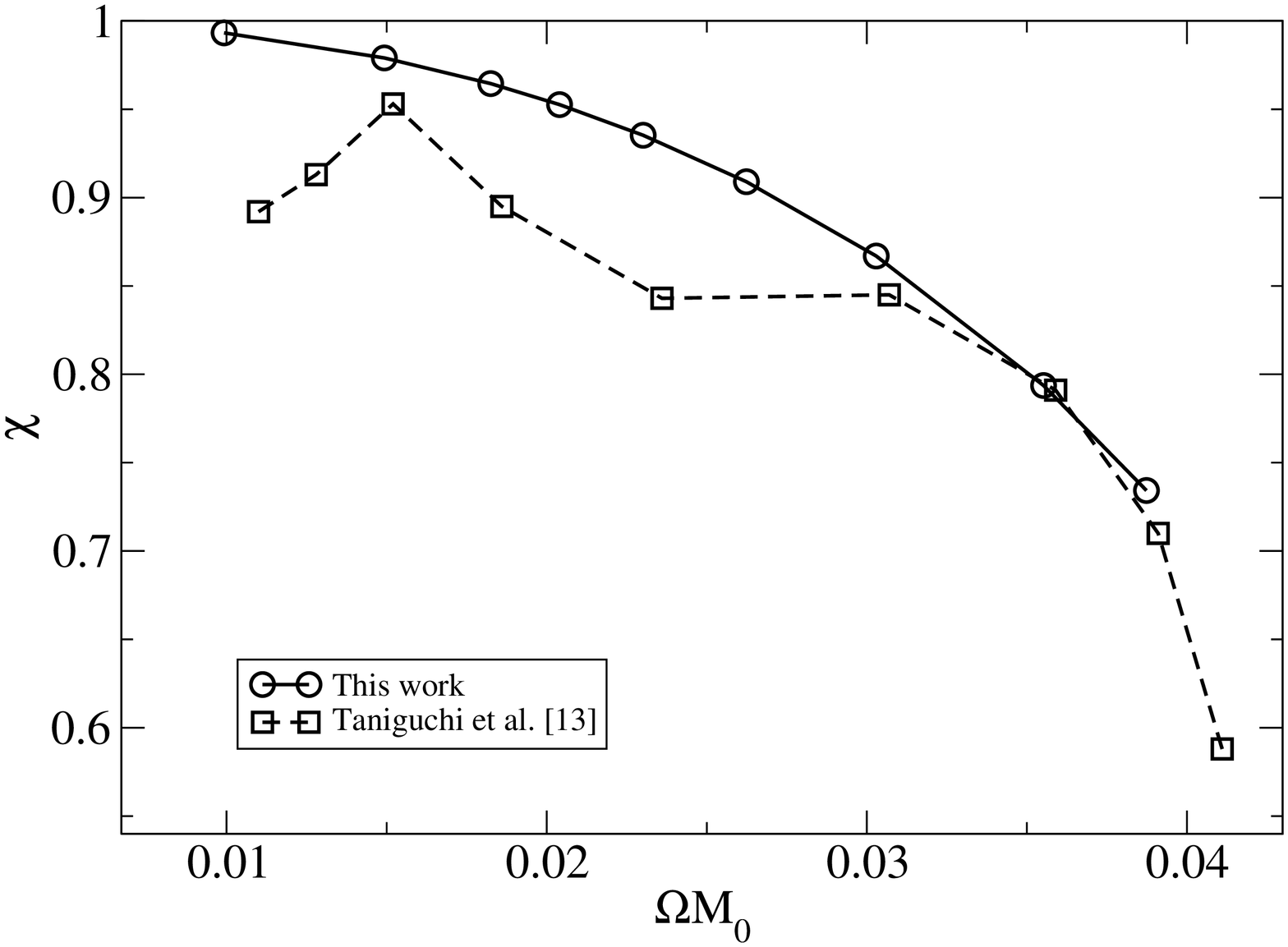}
\includegraphics[height=6.5cm]{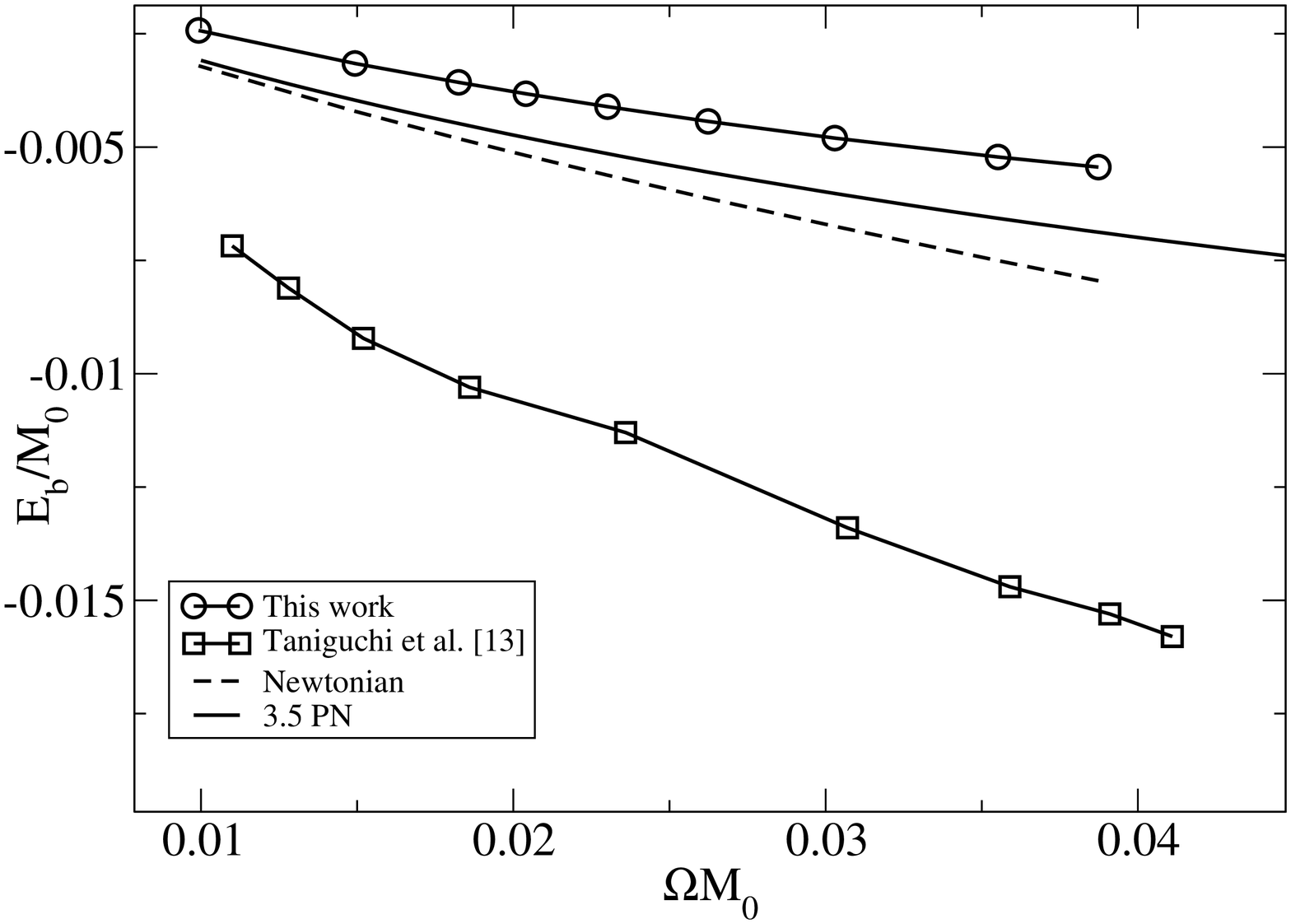}
\caption{\label{f:compare}
Comparison between this work and \cite{TanigBFS06} for a NS of small compactness ($\Xi=0.0879$). The ratio between the BH mass and the NS mass is 5.
The first panel shows the deformation parameter $\chi$ and the second one the binding energy, both as a function of the orbital velocity.}
\end{figure}

Another comparison can be made with the sequence presented in \cite{TanigBFS06} that relies on an approach similar to the work presented here. Let us first mention that the neutron star presented in \cite{TanigBFS06} is not realistic, having a compactness as small as $\Xi=0.0879$. For comparison purposes, a sequence with the same parameters has been computed with the code described here. In the first panel of Fig. \ref{f:compare}, the value of $\chi$ as a function of $\Omega M_0$ is shown. In all the following, $M_0$ denotes the total gravitational mass for infinite separation, which is the sum of the BH irreducible mass $M^{\rm irr}_{\rm BH}$ and the gravitational mass (i.e. the ADM mass) of the isolated NS: $M^{{\rm grav\, 0}}_{\rm NS}$. One of the most striking features of the first panel is that the data from \cite{TanigBFS06} behave in a different manner for large separations. Some of this surely comes from the Kerr-Schild approach, but the fact that the curve exhibits a very sharp maximum is puzzling. Moreover, for large separations, $\chi$ does not seem to converge to one as it should.
 On the contrary, the curve from this work exhibits a very smooth behavior and clearly goes to one for small values of the orbital frequency. For large values of $\Omega M_0$, that is, for small separations, the two curves agree reasonably well. The second panel of Fig. \ref{f:compare} shows the binding energy of the binary defined as $E_b=M_{\rm ADM}/M_0 -1$, as a function of $\Omega M_0$. The results from the two approaches are very different, the configurations from \cite{TanigBFS06} being more bounded by a factor of 3. For comparison, the binding energy from both Newtonian and 3.5 PN theories are shown (given by Eq. (194) of \cite{BlancLR}). The PN result is clearly much closer to the result of this paper. This closeness is a strong indication that the conformal flatness approximation is rather good because PN theory does not make us of it. The disagreement with 
\cite{TanigBFS06} may come from the differences in the choices of the ``freely-specifiable'' variables. If this is the case, it may indicate that the data generated with the Kerr-Shild approach have a spurious gravitational wave content. The total angular momenta coming from the two approaches are also compared and the agreement is good, the difference being only of the order of 
$4\%$ which is also the order of the difference with the PN result.

\section{More realistic neutron stars}
As already stated, the NS presented in the previous section has a very small compactness. In order to get more realistic neutron stars, this parameter should be increased. This is easily done by decreasing the parameter $\kappa$ in the EOS. The baryon mass of the NS is fixed so that they all have the same gravitational mass when isolated, i.e. the same $M^{{\rm grav\, 0}}_{\rm NS}$. Doing so, BHNS with four different compactness parameters (0.075, 0.100, 0.125 and 0.150) are constructed. As in the previous section, the ratio $M^{\rm irr}_{\rm BH}/M^{{\rm grav\, 0}}_{\rm NS}$ is set to 5.

\begin{figure}
\includegraphics[height=6.5cm]{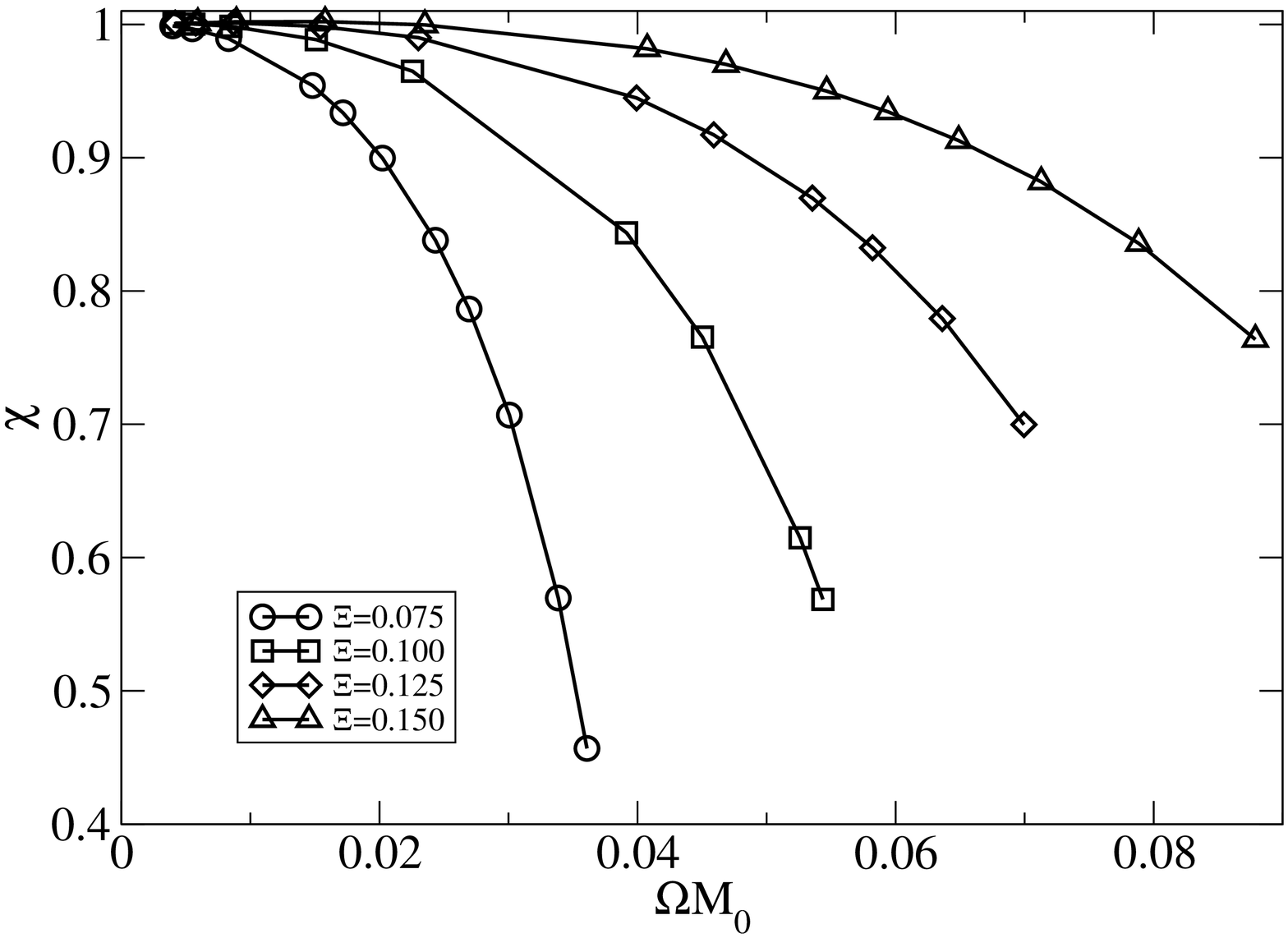}
\includegraphics[height=6.5cm]{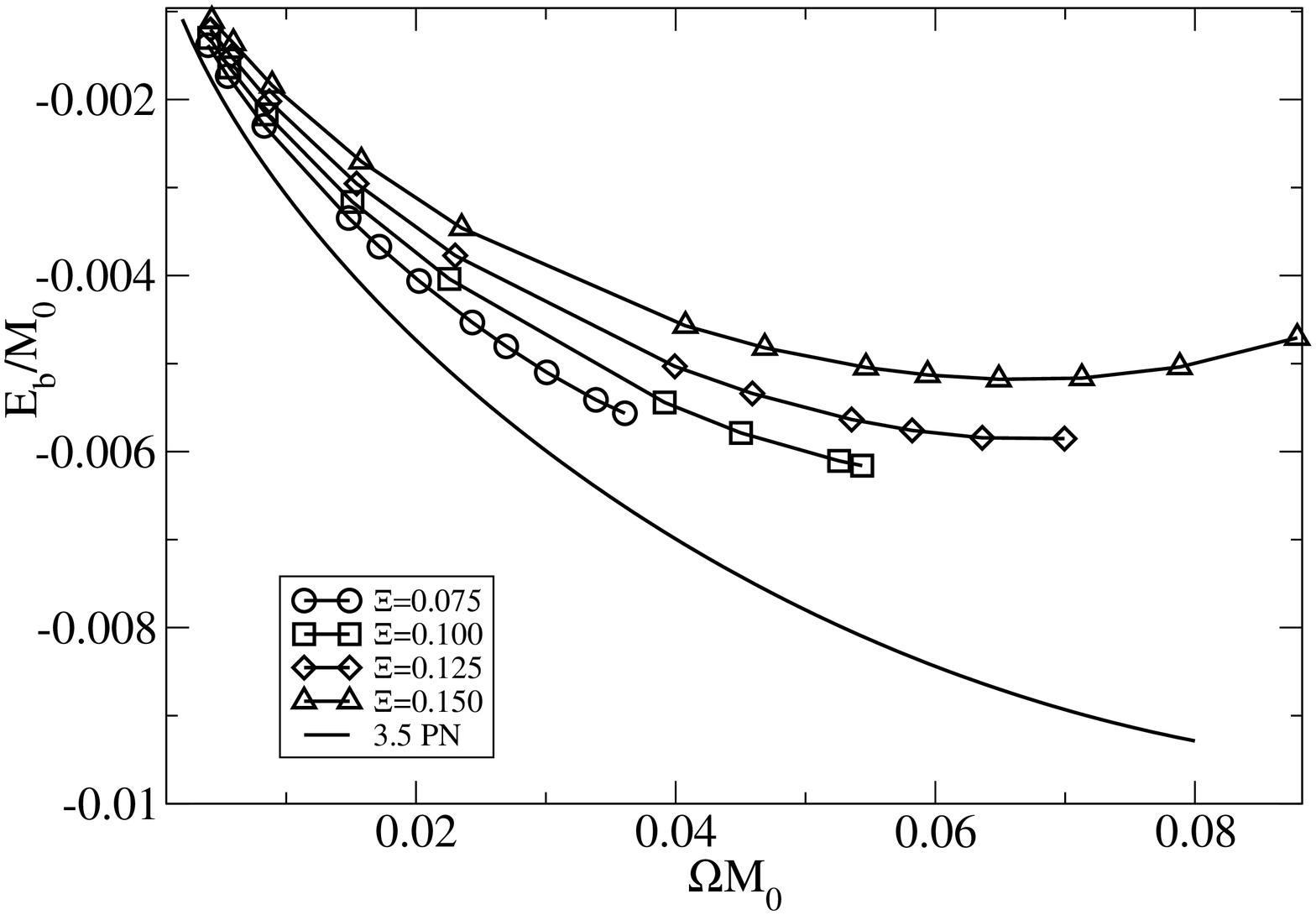}
\caption{\label{f:compact}
Deformation parameter $\chi$ (first panel) and binding energy (second panel) as a function of orbital velocity for four different compactness parameters $\Xi$. $M^{{\rm grav\, 0}}_{\rm NS}$ is the same for all the stars and the mass ratio with respect to the BH irreducible mass is 5.}
\end{figure}

\begin{figure}
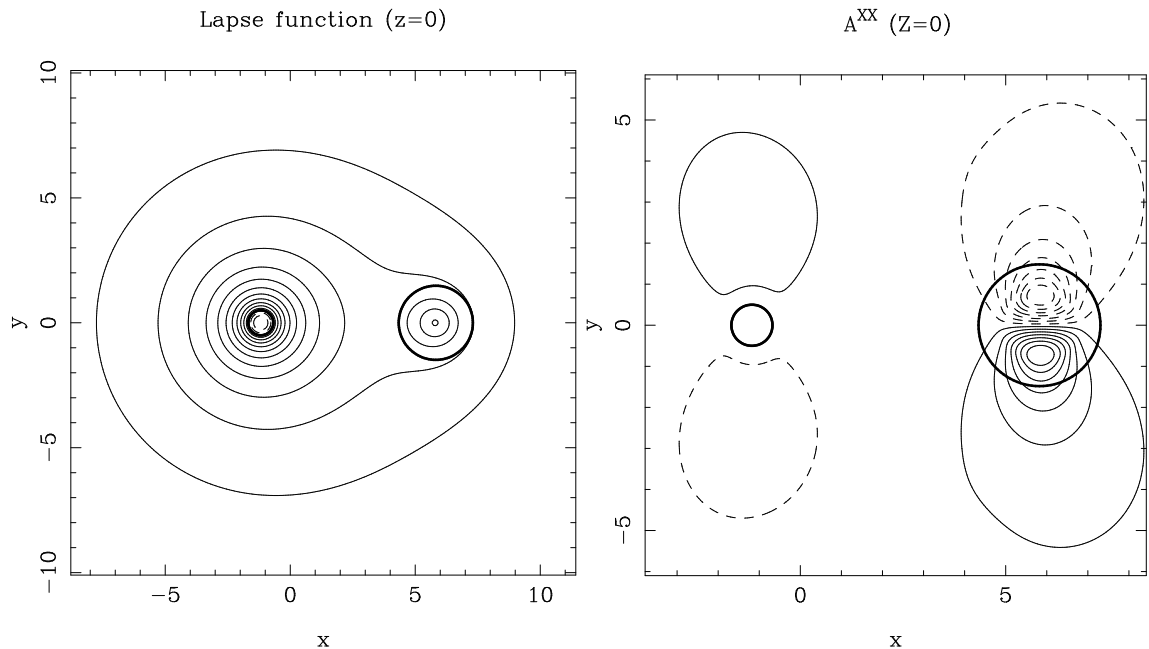

\includegraphics[height=8.5cm]{fig_4a.ps}
\includegraphics[height=8.5cm]{fig_4b.ps}
\caption{\label{f:fields}
Isocontours of the lapse (first panel) and of $\tilde{A}^{xx}$ (second panel) in the orbital plane. The thick black lines represent the surface of the compact objects (BH on the left and NS on the right). The distances are given in code coordinates.}
\end{figure}

The first panel of Fig. \ref{f:compact} shows the deformation $\chi$ as a function of $\Omega M_0$. As can be expected, the more compact stars are less easily deformed and can survive closer to the black hole without being tidally destroyed. The second panel shows the binding energy of the binaries, along with the 3.5 PN result for point masses. Once again, the more compact the NS, the closer it can get to the BH. It turns out that for the most compact star, the binding energy attains a minimum before the NS is destroyed. This may indicate that the system can reach dynamical instability, even if true stability can only be investigated by means of dynamical evolutions.
The minimum is also marginally attained for a compactness of 0.125. The two less compact stars, however, are destroyed before reaching the configuration of minimum binding energy. This explains why all the configurations of \cite{TanigBFS06} are found to be stable: the compactness of the star is small enough that it is destroyed before reaching the innermost stable orbit. The nature of the end point of a sequence thus depends on the compactness of the star, and this is an effect that could have some implications on the emitted gravitational signal. Fig. \ref{f:fields} shows the value of the lapse (first panel) and of $\tilde{A}^{xx}$ (second panel) in the orbital plane for a NS of compactness 0.15. The configuration is the one corresponding to the minimum of binding energy. The surfaces of both the NS (thick line on the right) and the BH (thick line on the left) are shown, and one can see that the NS is not yet noticeably deformed.

The total angular momentum $J$ can also be computed. When it admits a minimum (i.e. for the two most compact stars), its position in terms of frequency is consistent with the one for the binding energy (even if they do not coincide exactly). The relative importance of the BH local rotation can also be investigated by computing the ratio $f_r=\Omega_r/\Omega_0$. It turns out that $f_r$ is close to one (at most 0.91) which shows that the presence of the neutron star has a moderate influence on the structure of the BH horizon. $f_r$ is almost independent of the compactness of the star. However its dependence with frequency is different from Eq. (58) of \cite{CaudiCGP06}. This is probably simply an effect of the mass ratio (5 in this work and 1 in \cite{CaudiCGP06}). Finally the virial theorem states that the Komar-like mass and ADM mass should be equal for circular orbits. Contrary to the case of BBH \cite{GourgGB02,GrandGB02,CaudiCGP06}, this is not enforced exactly, the value of the orbital velocity being given by equilibrium of the NS fluid. However, the virial theorem can be used as a test of the code. It turns out that the virial theorem is verified to better than $2\%$ for all configurations, the dependency with the compactness of the NS being moderate. The difference between the two masses is greater for tighter configurations. This behavior is similar to what is found in \cite{TanigBFS06} even if the curves do not match, probably because of the different choice of ``freely specifiable'' variables. The violation of the virial theorem must be a measure on how the computed configurations differ from exact 
circularity. In a sense, it reflects the true nature of the movement: a slow inspiral.

All the configurations presented in this paper have been made public on the LORENE website \cite{lorene}.

\section{ERRATUM}
The version of the code used to compute the quasiequilibrium configurations presented in the previous sections was faulted by a wrong sign in some of the terms coupling the neutron star to the black hole. Indeed, terms like $\bar{D}_i N$ are split 
in a black hole and a neutron star part $\l. \bar{D}_i N \r|_{\rm BH} + \l. \bar{D}_i N\r|_{\rm NS}$. The terms generated by the neutron star had the wrong sign for the X and Y components, {\em when computed on the grid associated to the black hole}.

\begin{figure}
\includegraphics[height=6.5cm]{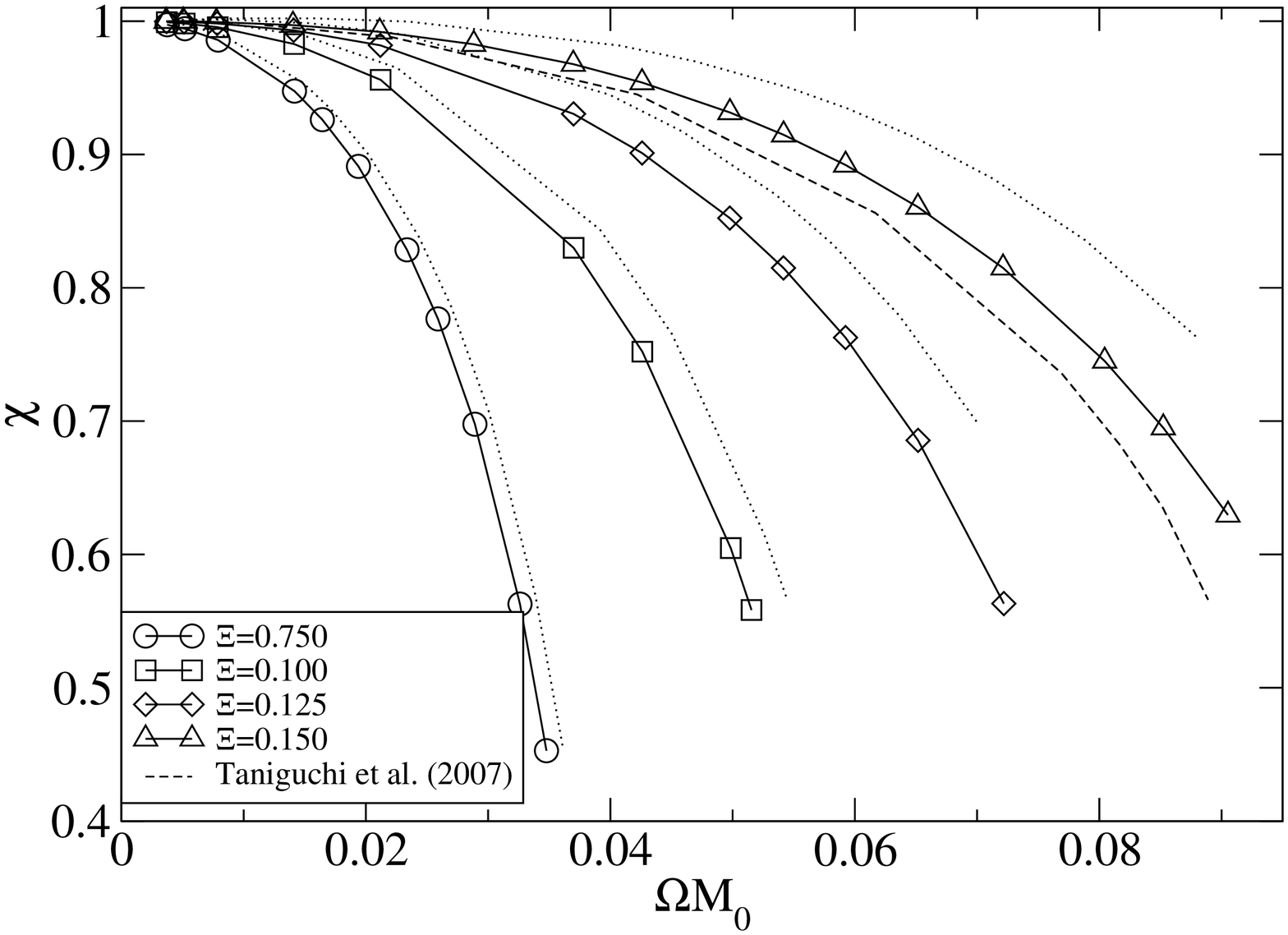}
\includegraphics[height=6.5cm]{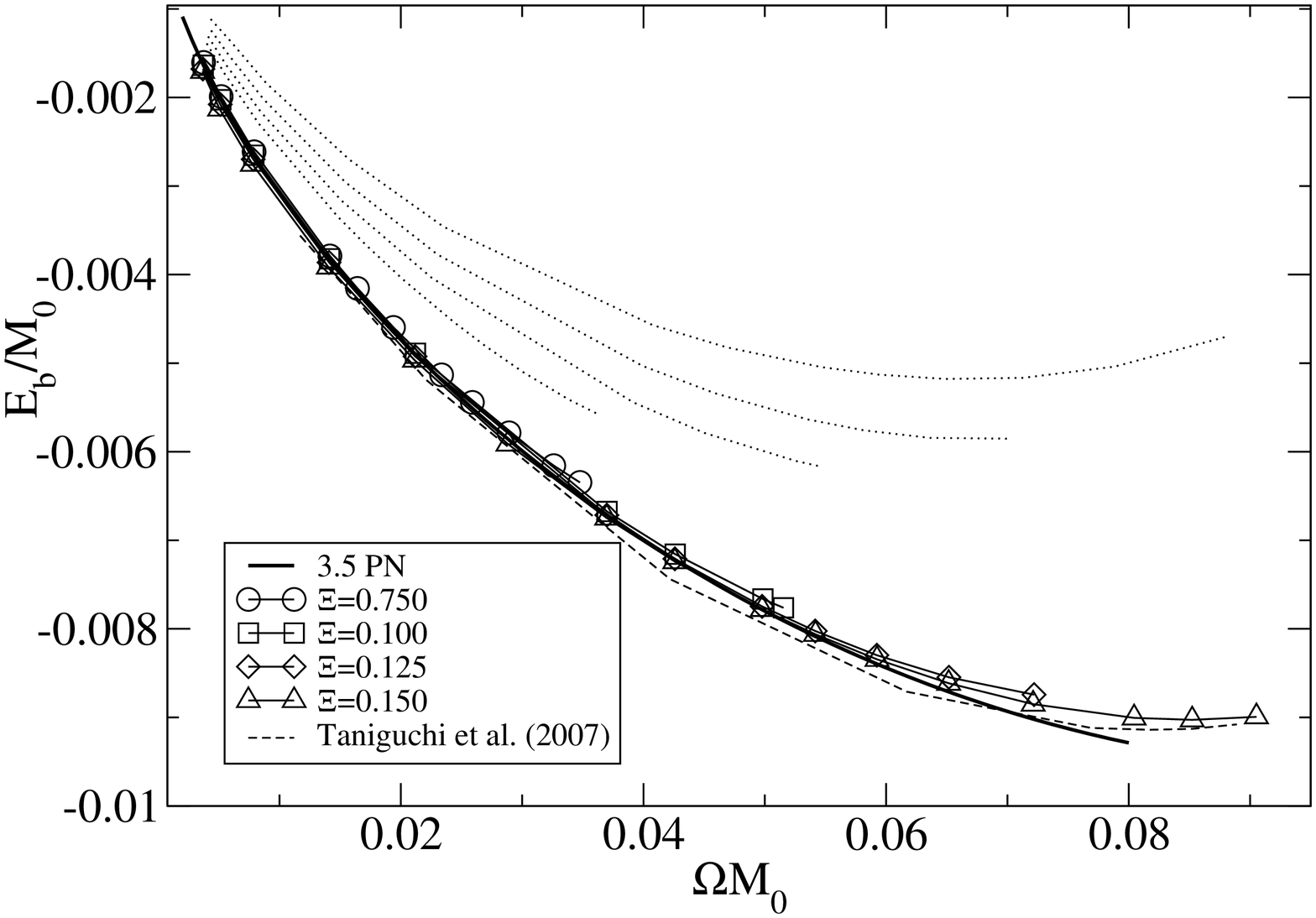}
\caption{\label{f:correct}
Correct version of Fig. \ref{f:compact}. \\
Deformation parameter $\chi$ (first panel) and binding energy (second panel) as a function of orbital velocity for four different compactness parameters $\Xi$. $M^{{\rm grav\, 0}}_{\rm NS}$ is the same for all the stars and the mass ratio with respect to the BH irreducible mass is 5. The dotted curves show the previous and not correct values and the dashed one the results from \cite{TanigBFS07}, for $\Xi=0.15$.}
\end{figure}

With the correct sign, the precision of the computation, as measured by the virial theorem, has improved by at least one order of magnitude. The obtained results are now in much better agreement with both post-Newtonian results and independent computations by Taniguchi {\em et al.} \cite{TanigBFS07}. This is clearly seen in Fig. \ref{f:correct} which is the correct version of Fig. 3. In particular, the more compact the star, the closer the binding energy is to the post-Newtonian curve, as expected. The new configurations can be accessed via the Lorene website 
\cite{lorene}.

\acknowledgments{
The author would like to express his gratitude to F.~Limousin for pointing out the mistake in the code.

\end{document}